\documentstyle[aps,preprint]{revtex}
\begin{document}
\author{Remo Garattini}
\address{Facolt\`{a} di Ingegneria, Universit\`{a} degli Studi di Bergamo,\\
Viale Marconi, 5, 24044 Dalmine (Bergamo) Italy\\
e-mail: Garattini@mi.infn.it}
\title{Casimir energy and variational methods in AdS spacetime}
\date{\today}
\maketitle

\begin{abstract}
Following the subtraction procedure for manifolds with boundaries, we
calculate by variational methods, the Schwarzschild-Anti-de Sitter and the
Anti-de Sitter space energy difference. By computing the one loop
approximation for TT tensors we discover the existence of an unstable mode
at zero temperature, which can be stabilized by the boundary reduction
method. Implications on a foam-like space are discussed.
\end{abstract}

\section{Introduction}

The problem of computing vacuum fluctuations in a field theory can be
considered as the first step to probe the validity of a theory. An example
is given by the zero point energy (ZPE) responsible of the Casimir effect.
This one was predicted by Casimir\cite{Casimir} and experimentally confirmed
in the Philips laboratories\cite{experiments}. This is induced when the
presence of electrical conductors distorts the zero point energy of the
quantum electrodynamics vacuum. Two parallel conducting surfaces, in a
vacuum environment, attract one another by a very weak force that varies
inversely as the fourth power of the distance between them. This kind of
energy is a pure quantum effect; no real particles are involved, only
virtual ones. The difference between the stress-energy computed in presence
and in absence of the plates with the same boundary conditions gives 
\begin{equation}
\Delta \left\langle T^{\mu \nu }\right\rangle =\left\langle T^{\mu \nu
}\right\rangle _{vac}^{p}-\left\langle T^{\mu \nu }\right\rangle _{vac}=%
\frac{\pi ^{2}}{720a^{4}}\left( 
\begin{array}{cccc}
-1 & 0 & 0 & 0 \\ 
0 & 1 & 0 & 0 \\ 
0 & 0 & 1 & 0 \\ 
0 & 0 & 0 & -3
\end{array}
\right) .
\end{equation}
It is evident that separately, each contribution coming from the summation
over all possible resonance frequencies of the cavities is divergent and
devoid of physical meaning but the {\it difference} between them in the two
situations (with and without the plates) is well defined. Note that the
energy density 
\begin{equation}
\rho =E/V=\Delta \left\langle T^{00}\right\rangle =-\frac{\pi ^{2}}{720a^{4}}
\end{equation}
is negative\cite{B.S. DeWitt,MVisser}. One can in general formally define
the Casimir energy as follows 
\begin{equation}
E_{Casimir}\left[ \partial {\cal M}\right] =E_{0}\left[ \partial {\cal M}%
\right] -E_{0}\left[ 0\right] ,
\end{equation}
where $E_{0}$ is the zero-point energy and $\partial {\cal M}$ is a
boundary. In General Relativity, at the classical level, there exists a
subtraction procedure related to the Arnowitt-Deser-Misner (ADM) approach 
\cite{ADM}, namely the ADM energy or mass, which can be improperly thought
as the classical aspect of the Casimir energy. In a recent paper, the
problem of computing the Casimir energy in presence of the Schwarzschild
metric for the gravitational field has been considered\cite{Remo}. The
classical energy associated to the related gravitational Casimir energy is
represented by the ADM\ mass 
\begin{equation}
M=\lim_{r\rightarrow \infty }\int_{\partial \Sigma }\sqrt{\widehat{g}}%
\widehat{g}^{ij}\left[ \widehat{g}_{ik,j}-\widehat{g}_{ij,k}\right] dS^{k},
\end{equation}
where $\widehat{g}^{ij}$ is the metric induced on a spacelike hypersurface $%
\partial \Sigma $ which has a boundary at infinity like $S^{2}$. An
equivalent definition of the classical energy is given by the quasilocal
energy defined by 
\begin{equation}
E_{q.l.}=\text{ }\frac{1}{8\pi G}\int_{S^{2}}d^{2}x\sqrt{\sigma }\left(
k-k^{0}\right) ,
\end{equation}
where $k$ is the extrinsic curvature referred to the Schwarzschild space and 
$k^{0}$ is the extrinsic curvature referred to flat space. $\sigma $ is the
two-dimensional determinant coming from the induced metric $\sigma _{ab}$ on
the boundaries $S^{2}$. It is relevant to observe that the Schwarzschild
space is asymptotically flat, namely when $r\rightarrow \infty $ we recover
the flat metric. In this case to correctly compute the classical energy term
a subtraction procedure is involved as widely discussed in Refs.\cite
{FroMar,HawHor}. When we transpose this procedure to one loop calculations,
we get the zero point energy (ZPE) for gravitons embedded in flat space 
\begin{equation}
2\cdot \frac{1}{2}\int \frac{d^{3}k}{\left( 2\pi \right) ^{3}}\sqrt{k^{2}}.
\end{equation}
This term has a quartic ultra-violet (UV)\ divergence. The same kind of
divergence is present when the Schwarzschild background is considered.
However their difference has a divergence degree of a logaritmic type. Since
boundary conditions are the same, this ZPE's difference at one loop
represents a Casimir-like computation. In this paper we would like to extend
the same evaluation reported in Ref.\cite{Remo,Remo1} to the computation of
the Casimir-like energy for a Schwarzschild-Anti-de Sitter (S-AdS) space at
zero temperature discussing the possible existence of an unstable mode. The
reason to compute such a correction comes from an analogy between the
Schwarzschild metric and the S-AdS metric. Indeed both metrics are
spherically symmetric and are characterized by only one root in the
gravitational potential. Moreover, no natural outer boundary is present. The
rest of the paper is structured as follows, in section \ref{p0} we define
the S-AdS line element, in section \ref{p1} we compute the quasilocal energy
and the quasilocal mass for the S-AdS space, in section \ref{p2} we give
some of the basic rules to perform the functional integration and we define
the Hamiltonian approximated up to second order, in section \ref{p3} we look
for stable modes of the spin-two operator acting on transverse traceless
tensors, in section \ref{p4} we show the existence of only one negative
mode, in section \ref{p5} we find a critical radius below which we have a
stabilization of the system. We summarize and conclude in section \ref{p6}.

\section{The Schwarzschild-Anti-de Sitter metric}

\label{p0}The S-AdS line element is defined as 
\begin{equation}
ds^{2}=-f\left( r\right) dt^{2}+f\left( r\right) ^{-1}dr^{2}+r^{2}d\Omega
^{2},  \label{i1}
\end{equation}
where 
\begin{equation}
f\left( r\right) =\left( 1-\frac{2MG}{r}+\frac{r^{2}}{b^{2}}\right) ,
\label{i1a}
\end{equation}
$b^{2}=\sqrt{-\frac{3}{\Lambda _{c}}}$ and $\Lambda _{c}<0$ is the negative
cosmological constant. For $\Lambda _{c}=0$ the metric describes the
Schwarzschild metric, while for $M=0$, we obtain 
\begin{equation}
ds^{2}=-\left( 1+\frac{r^{2}}{b^{2}}\right) dt^{2}+\left( 1+\frac{r^{2}}{%
b^{2}}\right) ^{-1}dr^{2}+r^{2}d\Omega ^{2},
\end{equation}
i.e. the Anti-de Sitter metric (AdS). The gravitational potential $f\left(
r\right) $ of $\left( \ref{i1}\right) $ has only one root located at 
\begin{equation}
\bar{r}=\sqrt[3]{\frac{3MG}{\Lambda _{c}}+\sqrt{\frac{1}{\Lambda _{c}^{2}}%
\left( 9\left( MG\right) ^{2}+\frac{1}{\Lambda _{c}}\right) }}+\sqrt[3]{%
\frac{3MG}{\Lambda _{c}}-\sqrt{\frac{1}{\Lambda _{c}^{2}}\left( 9\left(
MG\right) ^{2}+\frac{1}{\Lambda _{c}}\right) }}
\end{equation}
and the gravitational potential can be written as 
\begin{equation}
f\left( r\right) =\frac{\left( r-\bar{r}\right) \left( r^{2}+\bar{r}r+\bar{r}%
^{2}+b^{2}\right) }{rb^{2}}.
\end{equation}
From Eq.$\left( \ref{i1a}\right) $, evaluated at $\bar{r}$, the parameter $M$
can be written as 
\begin{equation}
MG=\frac{\bar{r}\left( \bar{r}^{2}+b^{2}\right) }{2b^{2}}.  \label{i1b}
\end{equation}
In complete analogy with the Schwarzschild case, we will consider a constant
time slice $\Sigma $ of the S-AdS manifold ${\cal M}$\footnote{%
In Appendix \ref{app1}, we will report the details concerning the
Kruskal-Szekeres description of the S-AdS manifold.}. Even if there is a
cosmological constant term we generalize the terminology by saying that the
hypersurface $\Sigma $ is an Einstein-Rosen bridge with wormhole topology $%
S^{2}\times R^{1}$. The Einstein-Rosen bridge defines a bifurcation surface
dividing $\Sigma $ in two parts denoted by $\Sigma _{+}$ and $\Sigma _{-}$.
Our purpose is to consider perturbations at $\Sigma $ with $t$ constant in
absence of matter fields, which naturally define quantum fluctuations of the
Einstein-Rosen bridge. The explicit expression of the Hamiltonian can be
calculated by means of the following line element 
\begin{equation}
ds^{2}=-N^{2}\left( dx^{0}\right) ^{2}+g_{ij}\left(
N^{i}dx^{0}+dx^{i}\right) \left( N^{j}dx^{0}+dx^{j}\right) ,  \label{i2}
\end{equation}
where $N$ is called the {\it lapse} function and $N_{i}$ is the {\it shift }%
function. When $N=\sqrt{1-\frac{2MG}{r}+\frac{r^{2}}{b^{2}}}$, $N_{i}=0$ and 
\begin{equation}
g_{ij}dx^{i}dx^{j}=\left( 1-\frac{2MG}{r}+\frac{r^{2}}{b^{2}}\right)
^{-1}dr^{2}+r^{2}d\Omega ^{2},
\end{equation}
we recover the S-AdS line element. On the slice $\Sigma $, deviations from
the S-AdS metric spatial section of the form 
\begin{equation}
g_{ij}=\bar{g}_{ij}+h_{ij}  \label{i3}
\end{equation}
will be considered with $N_{i}=0$ and $N\equiv N\left( r\right) $. Then the
line element $\left( \ref{i2}\right) $ becomes 
\begin{equation}
ds^{2}=-N^{2}\left( r\right) \left( dx^{0}\right) ^{2}+g_{ij}dx^{i}dx^{j}
\end{equation}
and the total Hamiltonian is 
\begin{equation}
H_{T}=H_{\Sigma }+H_{\partial \Sigma }=\int_{\Sigma }d^{3}x(N{\cal H+}N_{i}%
{\cal H}^{i})+H_{\partial \Sigma },
\end{equation}
where 
\begin{equation}
\left\{ 
\begin{array}{l}
{\cal H}{\bf =}G_{ijkl}\pi ^{ij}\pi ^{kl}\left( \frac{16\pi G}{\sqrt{g}}%
\right) -\left( \frac{\sqrt{g}}{16\pi G}\right) \left( R^{\left( 3\right) }+%
\frac{6}{b^{2}}\right) \ \text{ (Super Hamiltonian)} \\ 
{\cal H}^{i}=-2\pi _{|j}^{ij}\ \text{(Super Momentum)}
\end{array}
.\right.
\end{equation}
and $H_{\partial \Sigma }$ represents the energy stored into the boundary.
According to Witten\cite{Witten}, to discuss the existence of an unstable
sector, we have to compare spaces with the same boundary conditions. An
instability appears when we consider the Euclidean S-AdS spacetime with a
periodically identified time representing the equilibrium temperature of a
S-AdS black hole with the environment\cite{Prestidge}. The same boundary
conditions on the AdS spacetime can be imposed if the temperature on the
boundary is the same. Indeed the AdS spacetime has no natural temperature
and this seems to suggest that only the ``{\it hot}'' AdS spacetime will be
unstable. However, by applying the same method of Ref.\cite{Remo1}, it is
possible to discuss if the instability appears even when we have the $T=0$
temperature case. To this purpose the expression we need to evaluate is 
\begin{equation}
E^{S-AdS}\left( M,b\right) =E^{AdS}\left( b\right) +\Delta
E_{AdS}^{S-AdS}\left( M,b\right) _{|classical}+\Delta E_{AdS}^{S-AdS}\left(
M,b\right) _{|1-loop}.  \label{i4}
\end{equation}
$E^{AdS}\left( b\right) $ represents the reference space energy which is
zero for the flat space. $\Delta E_{AdS}^{S-AdS}\left( M,b\right)
_{|classical}$represents the energy difference stored in the boundaries due
to the presence of the hole and $\Delta E_{AdS}^{S-AdS}\left( M,b\right)
_{|1-loop}$ is the quantum correction to a classical term.

\section{Quasilocal Energy and Quasilocal Mass for the S-AdS space}

\label{p1}In this section we fix our attention on the classical part of Eq.$%
\left( \ref{i4}\right) $. We consider the outer boundary located at some
radius $R$. Thus the total energy at the classical level is 
\begin{equation}
E^{S-AdS}\left( M,b\right) =E^{AdS}\left( b\right) +\Delta
E_{AdS}^{S-AdS}\left( M,b\right) _{|classical}.
\end{equation}
We begin by looking at the ``{\it outside region''} of the Kruskal manifold
associated to the S-AdS spacetime\footnote{%
See Appendix \ref{app1} for details.}. We will use the quasilocal energy to
evaluate $\Delta E_{AdS}^{S-AdS}\left( M,b\right) _{|classical}$. Quasilocal
energy is defined as the value of the Hamiltonian that generates unit time
translations orthogonal to the two-dimensional boundary, 
\begin{equation}
\Delta E_{AdS}^{S-AdS}\left( M,b\right) _{|classical}=\frac{1}{8\pi G}%
\int_{S^{2}}d^{2}x\sqrt{\sigma }\left( k-k^{0}\right) ,
\end{equation}
where $\left| N\right| =1$ at $S^{2}$ and $k$ is the trace of the extrinsic
curvature corresponding to the S-AdS space and $k^{0}$ is the trace of the
extrinsic curvature referred to the AdS space. Following Refs. \cite
{BCM,FroMar,HawHor} and by means of Eq.$\left( \ref{i1a}\right) ,$ we obtain 
\[
\Delta E_{AdS}^{S-AdS}\left( M,b\right) _{|classical}=-\frac{1}{8\pi G}%
\int_{S^{2}}d\Omega ^{2}r^{2}\left[ \frac{-2\sqrt{f\left( r\right) }}{r}+%
\frac{2\sqrt{f\left( r\right) _{|M=0}}}{r}\right] _{|r=R} 
\]
\begin{equation}
=\frac{R}{G}\left[ \sqrt{1-\frac{2MG}{r}+\frac{r^{2}}{b^{2}}}-\sqrt{1+\frac{%
r^{2}}{b^{2}}}\right] 
%TCIMACRO{\underset{R\gg b}{\simeq }}%
%BeginExpansion
\mathrel{\mathop{\simeq }\limits_{R\gg b}}%
%EndExpansion
\frac{Mb}{R}.  \label{p1a}
\end{equation}
When the boundary is pushed to infinity $\Delta E_{AdS}^{S-AdS}\left(
M,b\right) _{|classical}\rightarrow 0$. The same happens to the temperature
defined by 
\begin{equation}
T=\left( \frac{\partial E}{\partial S}\right) _{r=R}=\frac{1}{2\pi }\frac{%
\kappa }{N\left( R\right) },
\end{equation}
where $N\left( R\right) =\sqrt{f\left( R\right) }$ is the redshift factor
and $\kappa $ is the surface gravity defined by 
\begin{equation}
\lim_{r\rightarrow \bar{r}}\frac{1}{2}\left| g_{00}^{^{\prime }}\left(
r\right) \right| =\frac{3\bar{r}^{2}+b^{2}}{2\bar{r}b^{2}}.  \label{p1aa}
\end{equation}
Thus except the limiting case of pushing the boundary to infinity, the
temperature $T$ and the classical energy $\Delta E_{AdS}^{S-AdS}\left(
M,b\right) _{|classical}$ do not vanish and therefore we cannot consider the
problem of searching for unstable modes at zero temperature. Nevertheless if
we look at the whole S-AdS manifold, the total classical energy can be
written as 
\[
E^{S-AdS}\left( M,b\right) =E^{AdS}\left( b\right) +E_{tot}\left( M,b\right) 
\]
\begin{equation}
=E^{AdS}\left( b\right) +\Delta E_{AdS}^{S-AdS}\left( M,b\right)
_{|classical}^{+}+\Delta E_{AdS}^{S-AdS}\left( M,b\right) _{|classical}^{-}
\end{equation}
with 
\[
\Delta E_{AdS}^{S-AdS}\left( M,b\right) _{|classical}^{+}=\frac{1}{8\pi G}%
\int_{S_{+}^{2}}d^{2}x\sqrt{\sigma }\left( k-k^{0}\right) , 
\]
\begin{equation}
\Delta E_{AdS}^{S-AdS}\left( M,b\right) _{|classical}^{-}=-\frac{1}{8\pi G}%
\int_{S_{-}^{2}}d^{2}x\sqrt{\sigma }\left( k-k^{0}\right) ,
\end{equation}
and $\left| N\right| =1$ at both $S_{+}^{2}$ and $S_{-}^{2}$. $E_{tot}\left(
M,b\right) $ is the quasilocal energy of a spacelike hypersurface $\Sigma
=\Sigma _{+}\cup \Sigma _{-}$ bounded by two boundaries $S_{+}^{2}$ and $%
S_{-}^{2}$ located in the two disconnected regions ${\cal M}_{+}$ and ${\cal %
M}_{-}$ respectively. To evaluate $\Delta E_{AdS}^{S-AdS}\left( M,b\right)
_{|classical}^{\pm }$ we can use Eq.$\left( \ref{p1a}\right) $ or more
pictorially by looking at the static Einstein-Rosen bridge associated to the
S-AdS space, whose metric is 
\begin{equation}
ds^{2}=-N^{2}\left( r\right) dt^{2}+g_{xx}dx^{2}+r^{2}\left( x\right)
d\Omega ^{2},  \label{p11}
\end{equation}
where $N$, $g_{xx}$, and $r$ are functions of the radial coordinate $x$
continuously defined on ${\cal M}$, with 
\begin{equation}
dx=\pm \frac{dr}{\sqrt{1-\frac{2MG}{r}+\frac{r^{2}}{b^{2}}}},  \label{p11a}
\end{equation}
where the plus sign is relative to $\Sigma _{+}$, while the minus sign is
related to $\Sigma _{-}$. If we make the identification $N^{2}=1-\frac{2MG}{r%
}+\frac{r^{2}}{b^{2}}$, the line element $\left( \ref{p11}\right) $ reduces
to the S-AdS metric written in another form. The boundaries $S_{\pm }^{2}$
are located at coordinate values $x=\bar{x}^{\pm }$. The normal to the
boundaries is $n^{\mu }=\left( h^{xx}\right) ^{\frac{1}{2}}\delta _{y}^{\mu
} $. By using the expression of the trace 
\begin{equation}
k=-\frac{1}{\sqrt{h}}\left( \sqrt{h}n^{\mu }\right) _{,\mu },
\end{equation}
we obtain 
\begin{equation}
k^{S-AdS}=\left\{ 
\begin{array}{c}
-2r,_{x}/r\qquad on\ \Sigma _{+} \\ 
2r,_{x}/r\qquad on\ \Sigma _{-}
\end{array}
\right. .
\end{equation}
Thus the computation of $E_{+}$ gives exactly the result of Eq.$\left( \ref
{p1a}\right) $. On the other hand the computation of $E_{-}$ gives the same
value but with the reversed sign. Thus one gets\footnote{%
Note that if we take as a reference space the flat space, then the trace is
taken to be $k^{flat}=-2/r$ and 
\[
\left( E^{S-AdS}-E^{flat}\right) _{\pm }=\left\{ 
\begin{array}{c}
-R^{2}/Gb\qquad S_{+}^{2} \\ 
R^{2}/Gb\qquad S_{-}^{2}
\end{array}
\right. . 
\]
When $R\rightarrow \infty $, $\left( E^{S-AdS}-E^{flat}\right) _{\pm
}\rightarrow -\infty $.} 
\begin{equation}
\left( E^{S-AdS}-E^{AdS}\right) _{\pm }=\left\{ 
\begin{array}{c}
Mb/R\qquad on\ S_{+}^{2} \\ 
-Mb/R\qquad on\ S_{-}^{2}
\end{array}
\right. ,  \label{p11b}
\end{equation}
where for $E_{-}$ we have used the conventions relative to $\Sigma _{-}$ and 
$S_{-}^{2}$. Therefore for every value of the boundary $R$, (provided we
take symmetric boundary conditions with respect to the bifurcation surface,
even for the limiting value $R\rightarrow \infty $), we have 
\begin{equation}
E^{S-AdS}\left( M,b\right) =E^{AdS}\left( b\right) +Mb/R-Mb/R=E^{AdS}\left(
b\right) ,
\end{equation}
namely the energy is conserved for every choice of the boundary location.

\section{Energy Density Calculation in Schr\"{o}dinger Representation}

\label{p2}In previous section we have fixed our attention to the classical
part of Eq.$\left( \ref{i4}\right) $. In this section, we apply the same
calculation scheme of Refs.\cite{Remo,Remo1} to compute one loop corrections
to the classical S-AdS term. Like the Schwarzschild case, there appear two
classical constraints for the Hamiltonian 
\begin{equation}
\left\{ 
\begin{array}{l}
{\cal H}\text{ }=0 \\ 
{\cal H}^{i}=0
\end{array}
\right. ,
\end{equation}
which are satisfied both by the S-AdS and AdS metric and two {\it quantum}
constraints 
\begin{equation}
\left\{ 
\begin{array}{l}
{\cal H}\tilde{\Psi}\text{ }=0 \\ 
{\cal H}^{i}\tilde{\Psi}=0
\end{array}
\right. .
\end{equation}
${\cal H}\tilde{\Psi}$ $=0$ is known as the {\it Wheeler-DeWitt} equation
(WDW). Nevertheless, our purpose is the computation of 
\begin{equation}
\Delta E_{AdS}^{S-AdS}\left( M,b\right) _{|1-loop}=\frac{\left\langle \Psi
\left| H_{\Sigma }^{S-AdS}-H_{\Sigma }^{AdS}\right| \Psi \right\rangle }{%
\left\langle \Psi |\Psi \right\rangle }  \label{p21}
\end{equation}
where $H_{\Sigma }^{S-AdS}$ and $H_{\Sigma }^{AdS}$ are the total
Hamiltonians referred to the S-AdS and AdS spacetimes respectively for the
volume term\cite{Remo} and $\Psi $ is a wave functional obtained following
the usual WKB expansion of the WDW solution. In this context, the
approximated wave functional will be substituted by a {\it trial wave
functional} of the gaussian form according to the variational approach we
shall use to evaluate $\Delta E_{AdS}^{S-AdS}\left( M,b\right) _{|1-loop}$.
Following the same procedure of Refs.\cite{Remo,Remo1}, we expand the
three-scalar curvature $\int d^{3}x\sqrt{g}\left( R^{\left( 3\right)
}+6/b^{2}\right) $ up to $o\left( h^{2}\right) $ and we get 
\[
\int_{\Sigma }d^{3}x\sqrt{\bar{g}}\left[ -\frac{1}{4}h\triangle h+\frac{1}{4}%
h^{li}\triangle h_{li}-\frac{1}{2}h^{ij}\nabla _{l}\nabla _{i}h_{j}^{l}+%
\frac{1}{2}h\nabla _{l}\nabla _{i}h^{li}-\frac{1}{2}h^{ij}R_{ia}h_{j}^{a}+%
\frac{1}{2}hR_{ij}h^{ij}\right] 
\]
\begin{equation}
-\int_{\Sigma }d^{3}x\sqrt{\bar{g}}\left[ \frac{1}{4}h^{li}\left(
6/b^{2}\right) h_{li}\right] .
\end{equation}
To explicitly make calculations, we need an orthogonal decomposition for
both $\pi _{ij\text{ }}$and $h_{ij}$ to disentangle gauge modes from
physical deformations. We define the inner product

\begin{equation}
\left\langle h,k\right\rangle :=\int_{\Sigma }\sqrt{g}G^{ijkl}h_{ij}\left(
x\right) k_{kl}\left( x\right) d^{3}x,
\end{equation}
by means of the inverse WDW metric $G_{ijkl}$, to have a metric on the space
of deformations, i.e. a quadratic form on the tangent space at h, with

\begin{equation}
\begin{array}{c}
G^{ijkl}=(g^{ik}g^{jl}+g^{il}g^{jk}-2g^{ij}g^{kl})\text{.}
\end{array}
\end{equation}
The inverse metric is defined on co-tangent space and it assumes the form

\begin{equation}
\left\langle p,q\right\rangle :=\int_{\Sigma }\sqrt{g}G_{ijkl}p^{ij}\left(
x\right) q^{kl}\left( x\right) d^{3}x\text{,}
\end{equation}
so that

\begin{equation}
G^{ijnm}G_{nmkl}=\frac{1}{2}\left( \delta _{k}^{i}\delta _{l}^{j}+\delta
_{l}^{i}\delta _{k}^{j}\right) .
\end{equation}
Note that in this scheme the ``inverse metric'' is actually the WDW metric
defined on phase space. The desired decomposition on the tangent space of
3-metric deformations\cite{BergerEbin,York} is:

\begin{equation}
h_{ij}=\frac{1}{3}hg_{ij}+\left( L\xi \right) _{ij}+h_{ij}^{\bot }
\label{p21a}
\end{equation}
where the operator $L$ maps $\xi _{i}$ into symmetric tracefree tensors

\begin{equation}
\left( L\xi \right) _{ij}=\nabla _{i}\xi _{j}+\nabla _{j}\xi _{i}-\frac{2}{3}%
g_{ij}\left( \nabla \cdot \xi \right) .
\end{equation}
Thus the inner product between three-geometries becomes 
\[
\left\langle h,h\right\rangle :=\int_{\Sigma }\sqrt{g}G^{ijkl}h_{ij}\left(
x\right) h_{kl}\left( x\right) d^{3}x= 
\]
\begin{equation}
\int_{\Sigma }\sqrt{g}\left[ -\frac{2}{3}h^{2}+\left( L\xi \right)
^{ij}\left( L\xi \right) _{ij}+h^{ij\bot }h_{ij}^{\bot }\right] .
\label{p21b}
\end{equation}
With the orthogonal decomposition in hand we can define the trial wave
functional 
\begin{equation}
\Psi \left[ h_{ij}\left( \overrightarrow{x}\right) \right] ={\cal N}\exp
\left\{ -\frac{1}{4l_{p}^{2}}\left[ \left\langle hK^{-1}h\right\rangle
_{x,y}^{\bot }+\left\langle \left( L\xi \right) K^{-1}\left( L\xi \right)
\right\rangle _{x,y}^{\Vert }+\left\langle hK^{-1}h\right\rangle
_{x,y}^{Trace}\right] \right\} ,
\end{equation}
where ${\cal N}$ is a normalization factor. Since we are interested only to
the perturbations of the physical degrees of freedom, we will fix our
attention only to the TT tensor sector reducing therefore the previous form
into 
\begin{equation}
\Psi \left[ h_{ij}\left( \overrightarrow{x}\right) \right] ={\cal N}\exp
\left\{ -\frac{1}{4l_{p}^{2}}\left\langle hK^{-1}h\right\rangle _{x,y}^{\bot
}\right\} .
\end{equation}
This restriction is motivated by the fact that if an instability appears
this will be in the physical sector referred to TT tensors, namely a non
conformal instability. This means that does not exist a gauge choice that
can eliminate negative modes. This choice seems also corroborated by the
action decomposition of Ref.\cite{GriKos}, where only TT tensors contribute
to the partition function\footnote{%
See also Ref.\cite{EKP} for another point of view.}. Therefore to calculate
the energy density, we need to know the action of some basic operators on $%
\Psi \left[ h_{ij}\right] $. The action of the operator $h_{ij}$ on $|\Psi
\rangle =\Psi \left[ h_{ij}\right] $ is realized by 
\begin{equation}
h_{ij}\left( x\right) |\Psi \rangle =h_{ij}\left( \overrightarrow{x}\right)
\Psi \left[ h_{ij}\right] .
\end{equation}
The action of the operator $\pi _{ij}$ on $|\Psi \rangle $, in general, is

\begin{equation}
\pi _{ij}\left( x\right) |\Psi \rangle =-i\frac{\delta }{\delta h_{ij}\left( 
\overrightarrow{x}\right) }\Psi \left[ h_{ij}\right] .
\end{equation}
The inner product is defined by the functional integration: 
\begin{equation}
\left\langle \Psi _{1}\mid \Psi _{2}\right\rangle =\int \left[ {\cal D}h_{ij}%
\right] \Psi _{1}^{\ast }\left\{ h_{ij}\right\} \Psi _{2}\left\{
h_{kl}\right\} ,
\end{equation}
and by applying previous functional integration rules, we obtain the
expression of the one-loop-like Hamiltonian form for TT (traceless and
transverseless) deformations 
\begin{equation}
H^{\bot }=\frac{1}{4l_{p}^{2}}\int_{{\cal M}}d^{3}x\sqrt{g}G^{ijkl}\left[
K^{-1\bot }\left( x,x\right) _{ijkl}+\left( \triangle _{2}\right)
_{j}^{a}K^{\bot }\left( x,x\right) _{iakl}\right] .  \label{p22}
\end{equation}
The propagator $K^{\bot }\left( x,x\right) _{iakl}$ comes from a functional
integration and it can be represented as 
\begin{equation}
K^{\bot }\left( \overrightarrow{x},\overrightarrow{y}\right)
_{iakl}:=\sum_{N}\frac{h_{ia}^{\bot }\left( \overrightarrow{x}\right)
h_{kl}^{\bot }\left( \overrightarrow{y}\right) }{2\lambda _{N}\left(
p\right) },
\end{equation}
where $h_{ia}^{\bot }\left( \overrightarrow{x}\right) $ are the
eigenfunctions of $\triangle _{2j}^{a}$ and $\lambda _{N}\left( p\right) $
are infinite variational parameters.

\section{The Schwarzschild-Anti-de Sitter Metric spin 2 operator and the
evaluation of the energy density}

\label{p3}

The Spin-two operator for the S-AdS metric is defined by 
\begin{equation}
\left( \triangle _{2}\right) _{j}^{a}:=-\triangle \delta
_{j}^{a}+2R_{j}^{a}+6/b^{2}\delta _{j}^{a}  \label{p3a}
\end{equation}
where $\triangle $ is the curved Laplacian (Laplace-Beltrami operator) on a
S-AdS background and $R_{j\text{ }}^{a}$ is the mixed Ricci tensor whose
components are: 
\begin{equation}
R_{i}^{a}=\left\{ -\frac{2MG}{r^{3}}-2/b^{2},\frac{MG}{r^{3}}-2/b^{2},\frac{%
MG}{r^{3}}-2/b^{2}\right\} .
\end{equation}
Note that the form of the mixed Ricci tensor for the S-AdS space is the same
of the mixed Ricci tensor computed in the Schwarzschild space, except for
the presence of the negative cosmological term. We are led to study the
following eigenvalue equation 
\begin{equation}
\left( -\triangle \delta _{j}^{a}+2R_{j}^{a}+6/b^{2}\delta _{j}^{a}\right)
h_{a}^{i}=E^{2}h_{j}^{i}  \label{p31}
\end{equation}
where $E^{2}$ is the eigenvalue of the corresponding equation. In doing so,
we follow Regge and Wheeler in analyzing the equation as modes of definite
frequency, angular momentum and parity\cite{RW}. The quantum number
corresponding to the projection of the angular momentum on the z-axis will
be set to zero. This choice will not alter the contribution to the total
energy since we are dealing with a spherical symmetric problem. In this
case, Regge-Wheeler decomposition shows that the even-parity
three-dimensional perturbation is

\begin{equation}
h_{ij}^{even}\left( r,\vartheta ,\phi \right) =diag\left[ H\left( r\right)
\left( 1-\frac{2MG}{r}+\frac{r^{2}}{b^{2}}\right) ^{-1},r^{2}K\left(
r\right) ,r^{2}\sin ^{2}\vartheta K\left( r\right) \right] Y_{l0}\left(
\vartheta ,\phi \right) .  \label{p32}
\end{equation}
Representation $\left( \ref{p32}\right) $ shows a gravitational perturbation
decoupling. For a generic value of the angular momentum $L$, one gets

\begin{equation}
\left\{ 
\begin{array}{c}
\left( -\triangle _{l}-\frac{4MG}{r^{3}}+\frac{2}{b^{2}}\right) H\left(
r\right) =E_{l}^{2}H\left( r\right) \\ 
\\ 
\left( -\triangle _{l}+\frac{2MG}{r^{3}}+\frac{2}{b^{2}}\right) K\left(
r\right) =E_{l}^{2}K\left( r\right) .
\end{array}
\right.  \label{p33}
\end{equation}
The Laplacian restricted to $\Sigma $ can be written as

\begin{equation}
\triangle _{l}=\left( 1-\frac{2MG}{r}+\frac{r^{2}}{b^{2}}\right) \frac{d^{2}%
}{dr^{2}}+\left( \frac{2r-3MG}{r^{2}}+3\frac{r}{b^{2}}\right) \frac{d}{dr}-%
\frac{l\left( l+1\right) }{r^{2}}.  \label{p33a}
\end{equation}
Defining reduced fields

\begin{equation}
H\left( r\right) =\frac{h\left( r\right) }{r};\qquad K\left( r\right) =\frac{%
k\left( r\right) }{r},
\end{equation}
and passing to the proper geodesic distance from the {\it throat} of the
bridge defined by Eq.$\left( \ref{p11a}\right) $, the system $\left( \ref
{p33}\right) $ becomes\footnote{%
The system does not change in form if we make the minus choice in Eq.$\left( 
\ref{p11a}\right) $.}

\begin{equation}
\left\{ 
\begin{array}{c}
-\frac{d^{2}}{dx^{2}}h\left( x\right) +\left( V^{-}\left( x\right) +\frac{3}{%
b^{2}}\right) h\left( x\right) =E_{l}^{2}h\left( x\right) \\ 
\\ 
-\frac{d^{2}}{dx^{2}}k\left( x\right) +\left( V^{+}\left( x\right) +\frac{3}{%
b^{2}}\right) k\left( x\right) =E_{l}^{2}k\left( x\right)
\end{array}
\right.  \label{p34}
\end{equation}
with 
\begin{equation}
V^{\mp }\left( x\right) =\frac{l\left( l+1\right) }{r^{2}\left( x\right) }%
\mp \frac{3MG}{r\left( x\right) ^{3}}.
\end{equation}
When $r\longrightarrow \infty $, $x\left( r\right) \simeq b\ln r$ and $%
V\left( x\right) \longrightarrow 0$. When $r\longrightarrow r_{0}$, $x\left(
r\right) \simeq 0$ and 
\begin{equation}
V^{\mp }\left( x\right) \longrightarrow \frac{l\left( l+1\right) }{r_{0}^{2}}%
\mp \frac{3MG}{r_{0}^{3}}=const,
\end{equation}
where $r_{0}$ satisfies the condition $r_{0}>\bar{r}$. The solution of $%
\left( \ref{p34}\right) $, in both cases (S-AdS and AdS one) is 
\begin{equation}
h\left( px\right) =k\left( px\right) =\sqrt{\frac{2}{\pi }}\sin \left(
px\right) .
\end{equation}
This choice is dictated by the requirement that 
\begin{equation}
h\left( x\right) ,k\left( x\right) \rightarrow 0\text{\qquad when\qquad }%
x\rightarrow 0\ \left( \text{alternatively }r\rightarrow \bar{r}\right) .
\end{equation}
Thus the propagator becomes 
\begin{equation}
K_{\pm }^{\bot }\left( x,y\right) =\frac{V}{2\pi ^{2}}\int_{0}^{\infty
}dpp^{2}\frac{\sin \left( px\right) }{r\left( x\right) }\frac{\sin \left(
py\right) }{r\left( y\right) }\frac{Y_{l0}\left( \vartheta ,\phi \right)
Y_{l^{\prime }0}\left( \vartheta ,\phi \right) }{\lambda _{\pm }\left(
p\right) }  \label{p35}
\end{equation}
$\lambda _{\pm }\left( p\right) $ is referred to the potential function $%
V^{\pm }\left( x\right) $. Substituting Eq.$\left( \ref{p35}\right) $ in Eq.$%
\left( \ref{p22}\right) $ one gets (after normalization in spin space and
after a rescaling of the fields in such a way as to absorb $l_{p}^{2}$) 
\begin{equation}
E\left( M,b,\lambda \right) =\frac{V}{8\pi ^{2}}\sum_{l=0}^{\infty
}\sum_{i=1}^{2}\int_{0}^{\infty }dpp^{2}\left[ \lambda _{i}\left( p\right) +%
\frac{E_{i}^{2}\left( p,M,b,l\right) }{\lambda _{i}\left( p\right) }\right]
\label{p36}
\end{equation}
where 
\begin{equation}
E_{1,2}^{2}\left( p,M,b,l\right) =p^{2}+\frac{l\left( l+1\right) }{r_{0}^{2}}%
\mp \frac{3MG}{r_{0}^{3}}+\frac{3}{b^{2}},
\end{equation}
$\lambda _{i}\left( p\right) $ are variational parameters corresponding to
the eigenvalues for a (graviton) spin-two particle in an external field and $%
V$ is the volume of the system. By minimizing $\left( \ref{p36}\right) $
with respect to $\lambda _{i}\left( p\right) $ one obtains $\overline{%
\lambda }_{i}\left( p\right) =\left[ E_{i}^{2}\left( p,M,b,l\right) \right]
^{\frac{1}{2}}$ and 
\begin{equation}
E\left( M,b,\overline{\lambda }\right) =\frac{V}{8\pi ^{2}}%
\sum_{l=0}^{\infty }\sum_{i=1}^{2}\int_{0}^{\infty }dpp^{2}2\sqrt{%
E_{i}^{2}\left( p,M,b,l\right) }\text{ }
\end{equation}
with 
\[
p^{2}+\frac{l\left( l+1\right) }{r_{0}^{2}}+\frac{3}{b^{2}}>\frac{3MG}{%
r_{0}^{3}}. 
\]
For the S-AdS background we get 
\begin{equation}
E\left( M,b\right) =\frac{V}{4\pi ^{2}}\sum_{l=0}^{\infty }\int_{0}^{\infty
}dpp^{2}\left( \sqrt{p^{2}+c_{-}^{2}}+\sqrt{p^{2}+c_{+}^{2}}\right)
\label{p37}
\end{equation}
where 
\[
c_{\mp }^{2}=\frac{l\left( l+1\right) }{r_{0}^{2}}\mp \frac{3MG}{r_{0}^{3}}+%
\frac{3}{b^{2}}, 
\]
while when we refer to the AdS space we put $M=0$ and $c^{2}=$ $\frac{%
l\left( l+1\right) }{r_{0}^{2}}+\frac{3}{b^{2}}$. Then 
\begin{equation}
E\left( b\right) =\frac{V}{4\pi ^{2}}\sum_{l=0}^{\infty }\int_{0}^{\infty
}dpp^{2}\left( 2\sqrt{p^{2}+c^{2}}\right)  \label{p38}
\end{equation}
Now, we are in position to compute the difference between $\left( \ref{p37}%
\right) $ and $\left( \ref{p38}\right) $. Since we are interested in the $UV$
limit, we have 
\[
\Delta E\left( M,b\right) =E\left( M,b\right) -E\left( b^{2}\right) 
\]
\[
=\frac{V}{4\pi ^{2}}\sum_{l=0}^{\infty }\int_{0}^{\infty }dpp^{2}\left[ 
\sqrt{p^{2}+c_{-}^{2}}+\sqrt{p^{2}+c_{+}^{2}}-2\sqrt{p^{2}+c^{2}}\right] 
\]
\begin{equation}
=\frac{V}{4\pi ^{2}}\sum_{l=0}^{\infty }\int_{0}^{\infty }dpp^{3}\left[ 
\sqrt{1+\left( \frac{c_{-}}{p}\right) ^{2}}+\sqrt{1+\left( \frac{c_{+}}{p}%
\right) ^{2}}-2\sqrt{1+\left( \frac{c}{p}\right) ^{2}}\right]
\end{equation}
and for $p^{2}>>c_{\mp }^{2},c^{2}$, we obtain 
\[
\frac{V}{4\pi ^{2}}\sum_{l=0}^{\infty }\int_{0}^{\infty }dpp^{3}\left[ 1+%
\frac{1}{2}\left( \frac{c_{-}}{p}\right) ^{2}-\frac{1}{8}\left( \frac{c_{-}}{%
p}\right) ^{4}+1+\frac{1}{2}\left( \frac{c_{+}}{p}\right) ^{2}-\frac{1}{8}%
\left( \frac{c_{+}}{p}\right) ^{4}\right. 
\]
\begin{equation}
\left. -2-\left( \frac{c}{p}\right) ^{2}+\frac{1}{4}\left( \frac{c}{p}%
\right) ^{4}\right] =-\frac{V}{2\pi ^{2}}\frac{c_{M}^{4}}{8}\int_{0}^{\infty
}\frac{dp}{p},
\end{equation}
where $c_{M}^{2}=3MG/r_{0}^{3}$. We will use a cut-off $\Lambda $ to keep
under control the $UV$ divergence 
\begin{equation}
\int_{0}^{\infty }\frac{dp}{p}\sim \int_{0}^{\frac{\Lambda }{c_{M}}}\frac{dx%
}{x}\sim \ln \left( \frac{\Lambda }{c_{M}}\right) ,
\end{equation}
where $\Lambda \leq m_{p}.$ Thus $\Delta E\left( M,b\right) $ for high
momenta becomes 
\begin{equation}
\Delta E\left( M,b\right) \sim -\frac{V}{2\pi ^{2}}\frac{c_{M}^{4}}{16}\ln
\left( \frac{\Lambda ^{2}}{c_{M}^{2}}\right) =-\frac{V}{32\pi ^{2}}\left( 
\frac{3MG}{r_{0}^{3}}\right) ^{2}\ln \left( \frac{r_{0}^{3}\Lambda ^{2}}{3MG}%
\right) .
\end{equation}
and Eq.$\left( \ref{i4}\right) $ to one loop is 
\begin{equation}
E^{S-AdS}\left( M,b\right) -E^{AdS}\left( b\right) =-\frac{V}{32\pi ^{2}}%
\left( \frac{3MG}{r_{0}^{3}}\right) ^{2}\ln \left( \frac{r_{0}^{3}\Lambda
^{2}}{3MG}\right) .  \label{p38a}
\end{equation}
Like the Schwarzschild case, we observe that 
\begin{equation}
\lim_{M\rightarrow 0}\lim_{r\rightarrow \bar{r}}\Delta E\left( M,b\right)
\neq \lim_{r\rightarrow \bar{r}}\lim_{M\rightarrow 0}\Delta E\left(
M,b\right) .
\end{equation}
This behavior seems to confirm that quantum effects come into play when we
try to reach the horizon. By means of Eq.$\left( \ref{i1b}\right) $, $\Delta
E\left( M,b\right) $ becomes 
\begin{equation}
\Delta E\left( \bar{r},b\right) =-\frac{V}{32\pi ^{2}}\left( 3\frac{\bar{r}%
\left( \bar{r}^{2}+b^{2}\right) }{2b^{2}r_{0}^{3}}\right) ^{2}\ln \left( 
\frac{2b^{2}r_{0}^{3}\Lambda ^{2}}{3\bar{r}\left( \bar{r}^{2}+b^{2}\right) }%
\right) .
\end{equation}
If we set $\bar{r}=b/\sqrt{3}=r_{m}$, which is the location of the minimum
of the surface gravity, we find that $\Delta E\left( \bar{r},b\right) $ is
reduced to 
\begin{equation}
\Delta E\left( b\right) =-\frac{V}{32\pi ^{2}}\left( \frac{2}{\sqrt{3}%
r_{0}^{3}}b\right) ^{2}\ln \left( \frac{\sqrt{3}r_{0}^{3}\Lambda ^{2}}{2b}%
\right) .  \label{p38b}
\end{equation}
Note that in the terminology of the black hole thermodynamics $r_{m}$
corresponds to the unique black hole solution whose temperature reaches its
minimum. To better appreciate the result obtained in Eq.$\left( \ref{p38a}%
\right) $, we define a scale variable $x=3MG/\left( r_{0}^{3}\Lambda
^{2}\right) $ in such a way that $\Delta E\left( M\right) $ can be cast in
the form 
\begin{equation}
\Delta E\left( x\right) =\frac{V}{32\pi ^{2}}\Lambda ^{4}x^{2}\ln x.
\end{equation}
A stationary point is reached for $x=0$, namely the AdS space and another
stationary point is in $x=e^{-\frac{1}{2}}.$ This last one represents a
minimum of $\Delta E\left( x\right) $. This means that there is a
probability that the spacetime without the hole will decay into a spacetime
with a hole (not black hole). To see if this is really possible, we have to
establish if there exist unstable modes. However in case of Eq.$\left( \ref
{p38b}\right) $ the claim that the stationary point $x=0$ represents the AdS
space is more delicate. Indeed this corresponds to the vanishing of the
parameter $b$, leading to a diverging negative cosmological constant,
saturating the whole space.

\section{Searching for negative modes}

\label{p4}

In this paragraph we look for negative modes of the eigenvalue equation $%
\left( \ref{p3a}\right) $. For this purpose we restrict the analysis to the
S wave. Indeed, in this state the centrifugal term is absent and this gives
the function $V\left( x\right) $ a potential well form, which is different
when the angular momentum $l\geq 1$. Moreover the potential well appears
only for the $H$ component, whose eigenvalue equation is 
\begin{equation}
\left( -\triangle -\frac{4MG}{r^{3}}+\frac{2}{b^{2}}\right) H\left( r\right)
=-E^{2}H\left( r\right) .  \label{p40}
\end{equation}
$\triangle $ is the operator $\triangle _{l}$ of Eq.$\left( \ref{p33a}%
\right) $ with $l=0$ and $E^{2}>0$. By defining the reduced field $h\left(
r\right) =H\left( r\right) r$, Eq.$\left( \ref{p40}\right) $ becomes 
\begin{equation}
-\frac{d}{dr}\left( \sqrt{1-\frac{2MG}{r}+\frac{r^{2}}{b^{2}}}\frac{dh}{dr}%
\right) +\left( \frac{-3MG}{r^{3}}+\tilde{E}^{2}\right) \frac{h}{\sqrt{1-%
\frac{2MG}{r}+\frac{r^{2}}{b^{2}}}}=0,  \label{p41}
\end{equation}
where $\tilde{E}^{2}=3/b^{2}+E^{2}$. By means of Eq.$\left( \ref{p11a}%
\right) $, one gets 
\[
-\frac{dx}{dr}\frac{d}{dx}\left( \sqrt{1-\frac{2MG}{r}+\frac{r^{2}}{b^{2}}}%
\frac{dh}{dx}\frac{dx}{dr}\right) +\left( -\frac{3MG}{r^{3}}+\tilde{E}%
^{2}\right) \frac{h}{\sqrt{1-\frac{2MG}{r}+\frac{r^{2}}{b^{2}}}} 
\]
\begin{equation}
=-\frac{d}{dx}\left( \frac{dh}{dx}\right) +\left( -\frac{3MG}{r^{3}}+\tilde{E%
}^{2}\right) h=0.  \label{p42}
\end{equation}
Near the throat 
\begin{equation}
x\left( r\right) \simeq \frac{\sqrt{2\bar{r}}}{\sqrt{\kappa }}\sqrt{\left( 
\frac{r}{\bar{r}}-1\right) },
\end{equation}
where $\kappa $ is the surface gravity. By defining the dimensionless
variable $\rho =\frac{r}{\bar{r}}$, we obtain $\rho =1+y^{2}$ where 
\begin{equation}
y=\sqrt{\kappa }x/\sqrt{2\bar{r}}=\tilde{\kappa}x.  \label{p43}
\end{equation}
Then Eq.$\left( \ref{p42}\right) $ becomes 
\[
-\frac{d}{dy}\left( \frac{dh}{dy}\right) \tilde{\kappa}^{2}+\left( -\frac{3MG%
}{\left( \bar{r}\right) ^{3}\rho ^{3}\left( y\right) }+\tilde{E}^{2}\right)
h 
\]
\begin{equation}
=-\frac{d^{2}h}{dy^{2}}+\left( -\frac{3MG}{\tilde{\kappa}^{2}\left( \bar{r}%
\right) ^{3}\left( 1+y^{2}\right) ^{3}}+\lambda \right) h=0,
\end{equation}
where $\lambda =\tilde{E}^{2}/\tilde{\kappa}^{2}$. Expanding the potential
around $y=0$, one gets 
\begin{equation}
-\frac{d^{2}h}{dy^{2}}+\left( -\frac{3MG}{\tilde{\kappa}^{2}\left( \bar{r}%
\right) ^{3}}\left( 1-3y^{2}\right) +\lambda \right) h
\end{equation}
\begin{equation}
=-\frac{d^{2}h}{dy^{2}}+\left( \omega ^{2}y^{2}-\frac{3MG}{\tilde{\kappa}%
^{2}\left( \bar{r}\right) ^{3}}+\lambda \right) h=0,
\end{equation}
where $\omega =\sqrt{9MG/\left( \tilde{\kappa}^{2}\left( \bar{r}\right)
^{3}\right) }$. In this approximation we have obtained the equation of a
quantum harmonic oscillator equation whose spectrum is $E_{n}=\hbar \omega
\left( n+\frac{1}{2}\right) $. Since we are using natural units, we set $%
\hbar =1$ and 
\begin{equation}
\lambda _{n}=3MG/\left( \tilde{\kappa}^{2}\left( \bar{r}\right) ^{3}\right) -%
\sqrt{9MG/\left( \tilde{\kappa}^{2}\left( \bar{r}\right) ^{3}\right) }\left(
n+\frac{1}{2}\right) .
\end{equation}
After some algebraic calculation, we obtain 
\begin{equation}
\lambda _{n}=6\sqrt{\frac{b^{2}+\bar{r}^{2}}{b^{2}+3\bar{r}^{2}}}\left( 
\sqrt{\frac{b^{2}+\bar{r}^{2}}{b^{2}+3\bar{r}^{2}}}-\frac{1}{\sqrt{2}}\left(
n+\frac{1}{2}\right) \right) ,
\end{equation}
where we have used the relation $\left( \ref{i1b}\right) $. We see that 
\begin{equation}
\lambda _{0}=6\sqrt{\frac{b^{2}+\bar{r}^{2}}{b^{2}+3\bar{r}^{2}}}\left( 
\sqrt{\frac{b^{2}+\bar{r}^{2}}{b^{2}+3\bar{r}^{2}}}-\frac{1}{2\sqrt{2}}%
\right) .
\end{equation}
Since the eigenvalue must be positive, the following inequality must hold 
\begin{equation}
\sqrt{\frac{b^{2}+\bar{r}^{2}}{b^{2}+3\bar{r}^{2}}}>\frac{1}{2\sqrt{2}}%
\Longrightarrow 7b^{2}+5\bar{r}^{2}>0,
\end{equation}
which is always verified. To proof that there is only one eigenvalue, we
look at the second eigenvalue 
\begin{equation}
\lambda _{1}=6\sqrt{\frac{b^{2}+\bar{r}^{2}}{b^{2}+3\bar{r}^{2}}}\left( 
\sqrt{\frac{b^{2}+\bar{r}^{2}}{b^{2}+3\bar{r}^{2}}}-\frac{3}{2\sqrt{2}}%
\right) .
\end{equation}
The inequality 
\begin{equation}
\sqrt{\frac{b^{2}+\bar{r}^{2}}{b^{2}+3\bar{r}^{2}}}>\frac{3}{2\sqrt{2}}%
\Longrightarrow b^{2}+19\bar{r}^{2}<0,
\end{equation}
which is never verified since $b$ and $\bar{r}$ are real quantities. Thus we
can conclude that there is only {\bf one eigenvalue} and according to
Coleman \cite{Coleman}, this is a signal of a transition from a false vacuum
to a true one. The same unstable mode appears also when we introduce a
temperature and we look at the thermodynamic stability of a S-AdS black hole
within isothermal cavities\cite{BCM,Prestidge,HawPage}. In terms of $E^{2}$,
the eigenvalue is 
\[
E^{2}=-3/b^{2}-3\frac{b^{2}+\bar{r}^{2}}{2b^{2}\bar{r}^{2}}+\frac{3}{4b^{2}%
\bar{r}^{2}}\sqrt{\left( b^{2}+\bar{r}^{2}\right) \left( b^{2}+3\bar{r}%
^{2}\right) } 
\]
\begin{equation}
=-3\frac{b^{2}+3\bar{r}^{2}}{2b^{2}\bar{r}^{2}}+\frac{3}{4b^{2}\bar{r}^{2}}%
\sqrt{\left( b^{2}+\bar{r}^{2}\right) \left( b^{2}+3\bar{r}^{2}\right) }.
\end{equation}

\section{Boundary Reduction and stability}

\label{p5}An equivalent approach to Eq.$\left( \ref{p41}\right) $ can be set
up by means of a variational procedure applied on a functional whose minimum
represents the solution of the problem. Let us define 
\[
J\left( h,E^{2}\right) =\frac{1}{2}\int\limits_{0}^{x\left( a\right) }dx%
\left[ \left( \frac{dh\left( x\right) }{dx}\right) ^{2}-\frac{3MG}{%
r^{3}\left( x\right) }h^{2}\left( x\right) \right] +\frac{\tilde{E}^{2}}{2}%
\int\limits_{0}^{x\left( a\right) }dxh^{2}\left( x\right) , 
\]
where $dx$ is given by Eq.$\left( \ref{p11a}\right) $. Eq.$\left( \ref{p41}%
\right) $ is equivalent to finding the minimum of 
\begin{equation}
\tilde{E}^{2}=\frac{\int\limits_{0}^{x\left( a\right) }dx\left[ \left( \frac{%
dh\left( x\right) }{dx}\right) ^{2}-\frac{3MG}{r^{3}\left( x\right) }%
h^{2}\left( x\right) \right] }{\int\limits_{0}^{x\left( a\right)
}dxh^{2}\left( x\right) }.  \label{p51}
\end{equation}
For future purposes, we use the boundary conditions 
\begin{equation}
h\left( x\left( a\right) \right) =0.
\end{equation}
When $r\rightarrow \infty \Longrightarrow x\rightarrow \infty $ and Eq.$%
\left( \ref{p42}\right) $ becomes 
\begin{equation}
-\frac{d^{2}h}{dx^{2}}+\tilde{E}^{2}h=0
\end{equation}
whose asymptotic solution is\footnote{%
Although the asymptotic behaviour is such that 
\[
x\left( r\right) =\pm b\ln r, 
\]
for practical purposes, we prefer to use the variable $x\left( r\right) $.} 
\begin{equation}
h\left( x\right) =A\exp \left( -\tilde{E}x\right) +B\exp \left( \tilde{E}%
x\right) .
\end{equation}
Since asymptotically $\exp \left( \tilde{E}x\right) $ diverges, we set $B=0$%
, then $h\left( x\right) $ becomes $A\exp \left( -\tilde{E}x\right) $. If we
change the variables in a dimensionless form like Eq.$\left( \ref{p43}%
\right) $, we get 
\begin{equation}
\mu =\frac{\tilde{E}^{2}}{\tilde{\kappa}^{2}}=\frac{\int\limits_{0}^{\bar{y}%
}dy\left[ \left( \frac{dh\left( y\right) }{dy}\right) ^{2}-\frac{3MG}{\bar{r}%
^{3}\tilde{\kappa}^{2}\rho ^{3}\left( y\right) }h^{2}\left( y\right) \right] 
}{\int\limits_{0}^{y\left( a\right) }dyh^{2}\left( y\right) }.  \label{p52}
\end{equation}
The asymptotic behaviour of $h\left( x\right) $ suggests to choose $h\left(
\lambda ,y\right) =\exp \left( -\lambda y\right) $ as a trial function, and
Eq.$\left( \ref{p52}\right) $ becomes 
\begin{equation}
\mu \left( \lambda \right) =\lambda ^{2}-\frac{3MG}{\bar{r}^{3}\tilde{\kappa}%
^{2}}\frac{\int\limits_{0}^{\bar{y}}\frac{dy}{\rho ^{3}\left( y\right) }\exp
\left( -2\lambda y\right) }{\frac{1-\exp \left( -2\lambda y\right) }{%
2\lambda }}.
\end{equation}
Close to the throat $\exp \left( -2\lambda y\right) \simeq 1-2\lambda y$ and 
\begin{equation}
\mu \left( \lambda \right) =\lambda ^{2}-\frac{3MG}{\bar{r}^{3}\tilde{\kappa}%
^{2}}+\frac{9MG}{\bar{r}^{3}\tilde{\kappa}^{2}}\left[ \frac{\bar{y}}{%
2\lambda }+\bar{y}^{2}\right] .
\end{equation}
The minimum of $\mu \left( \lambda \right) $ is reached for $\bar{\lambda}%
=\left( \frac{9MG}{4\bar{r}^{3}\tilde{\kappa}^{2}}\bar{y}\right) ^{\frac{1}{3%
}}$ with the help of Eq.$\left( \ref{i1b}\right) $, we can write 
\begin{equation}
\mu \left( \bar{\lambda}\right) =3\left( \frac{9}{4}D\bar{y}\right) ^{\frac{2%
}{3}}-3D+3D\bar{y}^{2},  \label{p53}
\end{equation}
where 
\begin{equation}
D=\frac{MG}{\bar{r}^{3}\tilde{\kappa}^{2}}=2\frac{\bar{r}^{2}+b^{2}}{3\bar{r}%
^{2}+b^{2}}.
\end{equation}
If we set the value of $\bar{r}$ equal to the surface gravity minimum
location, then Eq.$\left( \ref{p53}\right) $becomes 
\begin{equation}
\mu \left( \bar{\lambda}\right) =3\left( 3\bar{y}\right) ^{\frac{2}{3}}-4+12%
\bar{y}^{2},
\end{equation}
which is zero for $\bar{y}_{c}=.309\,15$ corresponding to $\bar{\rho}%
_{c}=1.0956$. This means that the unstable mode persists until the boundary
radius $\bar{\rho}$ falls below $\bar{\rho}_{c}$, to be compared with the
value $\rho =1$ of Refs.\cite{Prestidge,HawPage}.

\section{Summary and Conclusions}

\label{p6}Following Refs.\cite{Remo,Remo1}, in this paper we have computed
the Casimir-like energy for a S-AdS space with a AdS space as a reference
space. Due to the same asymptotic properties of these spaces and looking at
the extended Kruskal S-AdS manifold at constant time, we have found that the
classical contribution coming from boundaries disappears. According to
Witten \cite{Witten}, since the energy is conserved and since boundary
conditions are the same we can discuss the existence of an instability at
zero temperature. A proof of instability at finite temperature has been
given by Prestidge in Ref.\cite{Prestidge}, based on conjectures of Hawking
and Page \cite{HawPage}. The zero temperature one-loop analysis shows a
single negative mode in S wave which is interpreted as a clear signal of a
decay form a false vacuum to a true one\cite{Coleman}. This is also
confirmed by one-loop stable modes contribution which shifts the energy
minimum to the S-AdS space. Following the same procedure of Ref.\cite{Remo2}%
, we discover a critical radius $r_{c}$ below which the system becomes
stable. In analogy with the flat space case, the appearance of this
instability even at zero temperature is attributed to a neutral S-AdS black
hole pair creation mediated by a three-dimensional S-AdS wormhole with the
holes residing in different universes\cite{Remo1}. The probability of
creating such a pair and therefore the instability appearance at zero
temperature is measured by 
\begin{equation}
\Gamma _{{\rm 1-hole}}=\frac{P_{{\rm S-AdS}}}{P_{{\rm AdS}}},
\end{equation}
where 
\begin{equation}
P\sim \left| \exp -\left( \Delta E\right) \left( \Delta t\right) \right|
^{2}.
\end{equation}
In spite of the evident analogy between the S-AdS and the Schwarzschild
space it is not simple at this stage speculate on a possible foam-like
structure composed by $N$ S-AdS coherent wormholes because the boundary
reduction, which is fundamental to have the stabilization of the system in
examination, is not of straightforward application due to the presence of
the negative cosmological factor proportional to the square of the radius.
However, if such a reduction mechanism could work $we$should discuss what is
the meaning of 
\begin{equation}
\Gamma _{{\rm N-S-AdS\ holes}}=\frac{P_{{\rm N-S-AdS\ holes}}}{P_{{\rm AdS}}}%
\simeq \frac{P_{{\rm foam}}}{P_{{\rm AdS}}}
\end{equation}
compared with 
\begin{equation}
\Gamma _{{\rm N-holes}}=\frac{P_{{\rm N-holes}}}{P_{{\rm flat}}}\simeq \frac{%
P_{{\rm foam}}}{P_{{\rm flat}}}.
\end{equation}

\section{Acknowledgements}

The author would like to thank the Referees for useful comments and
suggestions, in particular for having brought to my attention the paper of
Ref.\cite{GriKos}.

\appendix

\section{Kruskal-Szekeres coordinates for S-AdS spacetime}

\label{app1}

Before introducing the Kruskal-Szekeres\cite{KS,HawEll,MTW} type coordinates
we recall that the S-AdS line element is defined as 
\begin{equation}
ds^{2}=-\left( 1-\frac{2MG}{r}+\frac{r^{2}}{b^{2}}\right) dt^{2}+\left( 1-%
\frac{2MG}{r}+\frac{r^{2}}{b^{2}}\right) ^{-1}dr^{2}+r^{2}d\Omega ^{2}.
\end{equation}
By looking at the $\left( t,r\right) $ coordinates one gets 
\[
ds^{2}=-\left( 1-\frac{2MG}{r}+\frac{r^{2}}{b^{2}}\right) \left[
dt^{2}-dr^{\ast 2}\right] +r^{2}d\Omega ^{2} 
\]
\begin{equation}
=-\left( 1-\frac{2MG}{r}+\frac{r^{2}}{b^{2}}\right) dvdu+r^{2}\left(
u,v\right) d\Omega ^{2}.  \label{a1}
\end{equation}
$v=t+r^{\ast }$ is the ingoing radial null coordinate, $u=t-r^{\ast }$ is
the outgoing radial null coordinate and 
\begin{equation}
dr^{\ast }=\frac{rb^{2}dr}{\left( r-\bar{r}\right) \left( r^{2}+\bar{r}r+%
\bar{r}^{2}+b^{2}\right) }.
\end{equation}
The explicit integration gives $2\kappa r^{\ast }$%
\begin{equation}
=\ln \left| \frac{r}{\bar{r}}-1\right| -\frac{1}{2}\ln \left( \frac{r^{2}+%
\bar{r}r+\bar{r}^{2}+b^{2}}{\bar{r}^{2}+b^{2}}\right) +\frac{3\bar{r}%
^{2}+2b^{2}}{\bar{r}\sqrt{3\bar{r}^{2}+4b^{2}}}\arctan \left( \frac{2r\sqrt{3%
\bar{r}^{2}+4b^{2}}}{4\bar{r}^{2}+4b^{2}+\bar{r}r}\right) ,
\end{equation}
where $\kappa $ is the surface gravity defined in Eq.$\left( \ref{p38b}%
\right) $. We now introduce Kruskal-Szekeres coordinates $\left( U,V\right) $
defined (for $r>\bar{r}$) by 
\begin{equation}
U=-e^{-\kappa u}\qquad V=e^{\kappa v}
\end{equation}
with 
\begin{equation}
UV=-\exp \kappa \left( v-u\right) =-\exp \left( 2\kappa r^{\ast }\right)
\end{equation}
\begin{equation}
=-\frac{\left( r/\bar{r}-1\right) \sqrt{\bar{r}^{2}+b^{2}}}{\sqrt{r^{2}+\bar{%
r}r+\bar{r}^{2}+b^{2}}}\exp \left( \frac{3\bar{r}^{2}+2b^{2}}{\bar{r}\sqrt{3%
\bar{r}^{2}+4b^{2}}}\arctan \left( \frac{2r\sqrt{3\bar{r}^{2}+4b^{2}}}{4\bar{%
r}^{2}+4b^{2}+\bar{r}r}\right) \right)
\end{equation}
and 
\begin{equation}
\frac{U}{V}=-\exp -\kappa \left( v+u\right) =-\exp \left( -2\kappa t\right)
\end{equation}

In terms of these coordinates Eq.$\left( \ref{a1}\right) $ becomes 
\begin{equation}
ds^{2}=-\frac{\bar{r}\left( r^{2}+\bar{r}r+\bar{r}^{2}+b^{2}\right) ^{\frac{3%
}{2}}}{\sqrt{\bar{r}^{2}+b^{2}}rb^{2}\kappa ^{2}}\exp \left( F\left(
r\right) \right) dUdV+r^{2}\left( U,V\right) d\Omega ^{2},
\end{equation}
where 
\begin{equation}
F\left( r\right) =-\frac{3\bar{r}^{2}+2b^{2}}{\bar{r}\sqrt{3\bar{r}%
^{2}+4b^{2}}}\arctan \left( \frac{2r\sqrt{3\bar{r}^{2}+4b^{2}}}{4\bar{r}%
^{2}+4b^{2}+\bar{r}r}\right) .
\end{equation}
The only true singularities are at curves $UV=1$, where $r=0$. The region $%
\left\{ U<0,V>0\right\} $ is the ``{\it outside region}'', the only region
from which distant observers can obtain any information. The line $V=0$,
where $r=\bar{r}$, is the ``{\it past horizon}''; the line $U=0$ where also $%
r=\bar{r}$, is the ``{\it future horizon}''.

\end{document}